# Identifying Mitigation Strategies for COVID-19 Superspreading on Flights using Models that Account for Passenger Movement

Sirish Namilae[1], Yuxuan Wu[1], Anuj Mubayi[2], Ashok Srinivasan[3*], Matthew Scotch[4]

[1] Embry-Riddle Aeronautical University, Daytona Beach, Florida, USA
[2] PRECISIONheor, Los Angeles, California, USA
[3] University of West Florida, Pensacola, Florida, USA
[4] Arizona State University, Tempe, Arizona, USA
* Corresponding author: asrinivasan@uwf.edu

**Abstract:** Despite commercial airlines mandating masks, there have been multiple documented events of COVID-19 superspreading on flights. We used available data from three flights, including cabin layout and seat locations of infected and uninfected passengers, to suggest interventions to mitigate COVID-19 superspreading events during air travel. Specifically, we studied: 1) London to Hanoi with 201 passengers, including 13 secondary infections among passengers; 2) Singapore to Hangzhou with 321 passengers, including 12 to 14 secondary infections; 3) a non-superspreading event on a private jet in Japan with 9 passengers and no secondary infections. We show that inclusion of passenger movement better explains the infection spread patterns than conventional models do. We also found that FFP2/N95 mask usage would have reduced infection by 95-100%, while cloth masks would have reduced it by only 40-80%. This suggests the need for more stringent guidelines to reduce aviation-related superspreading events of COVID-19.

## Introduction

COVID-19 transmission is primarily driven by proximity between an infective person and a susceptible person [1]. Superspreading events, which involve the secondary infection of an unusually large number of persons [2], often occur when groups of people are brought into close proximity. However, crowded events do not necessarily lead to superspreading. While it is difficult to predict if an event would lead to superspreading, it is possible to take preventive steps to reduce its possibility.

There has been much concern about superspreading in airplanes, because large numbers of passengers are brought into close proximity there. While a lack of contact tracing has limited the availability of data on the extent of superspreading in planes, there have been multiple incidents recorded [3]. The Centers for Disease Control and Prevention (CDC) has, therefore, suggested guidelines, such as the use of masks, to reduce the risk of COVID-19 spread on planes [4].

Mathematical models can provide insight into the mechanisms of superspreading on planes and help evaluate the potential effectiveness of mitigation measures in reducing the likelihood of such events. However, conventional models are unable to adequately explain superspreading patterns on flights, with infection spread being wider than would be expected from proximity based on passenger seating [5, 6]. An important reason for this is that models typically do not consider the movement of passengers during the flight, boarding, or deplaning [7]. Understanding the risks for each of these aspects could provide insight into effective mitigation measures.

We have previously proposed an approach using pedestrian dynamics, a technique used to simulate the movement of individuals [8-10], to identify infection risk arising from proximity during boarding [11-14]. Here, we augment it with modeling of inflight transmission. We also use a new infection spread model that accounts for varying infection dose by distance to an infective person, and then include it in a standard exponential dose-response relationship for infection risk. It is difficult to identify model parameters *a priori*. Instead, we calibrate the model against a different superspreading event and modify the model to account for behavioral features such as mask wearing.

In this paper we: (1) explain the modeling methodology, which could be adopted in a wide variety of contexts; (2) quantify the role of different categories of passenger movements on infection transmission in airplanes; and (3) identify the impact of mask type on reducing the likelihood of superspreading events.

We show that our modeling approach can explain the wider spread of COVID-19 than expected in the superspreading examples considered. Our model also shows that N95 masks would be around ten time more effective than regular cloth masks in reducing superspreading.

*Flights studied*

We studied three flights that had detailed information on in-flight COVID-19 seating arrangements and infection profiles of the passengers as shown in Figure 1. A London flight is used to study the impact of passenger movement on infection risk when masks are not used. A Singapore flight examines the impact of mask wearing. We use a non-superspreading event of a Japan flight to validate the insight from our simulations that widespread use of N95 masks can greatly reduce the risk of infection.

The *London* flight departed from London to Hanoi on March 1st, 2020 [5]. The 10-hour flight had 16 crewmembers and 201 passengers onboard. Twenty-one of these passengers were in the business class cabin, 35 in premium economy, and 148 passengers were seated in the economy cabin. One index passenger was located in the business class, resulting in 11 secondary infections in the first-class and two in the economy cabin. Mask usage was not mandated on this flight, and its use was only sporadic.

The *Singapore* flight departed Singapore on January 24th, 2020 and landed in Hangzhou, China on January 25th [6, 15]. The total flight duration was five hours, with 321 passengers on board. Mask usage was mandatory for this flight. All infected cases were wearing the masks, although the mask type is not known [6, 15]. There were two index cases on the plane, sitting far apart, with one sitting next to the window in the aft-economy cabin and the other next to the aisle in the mid-economy cabin. An additional 14 passengers tested positive within the 14 days quarantine. Some of these passengers could have been exposed outside of the flight [6], and the number of inflight infections is estimated as 12 [15].

The *Japan* flight, a private jet, flew for 13.5 hours to Israel on Feb 20th, 2020 with 9 passengers, two of whom were infectious [16]. High quality FFP2 masks were used on this flight. No secondary infections occurred. We included this non-superspreading flight to increase the robustness of our model by examining a "control" scenario where secondary infections did not occur.

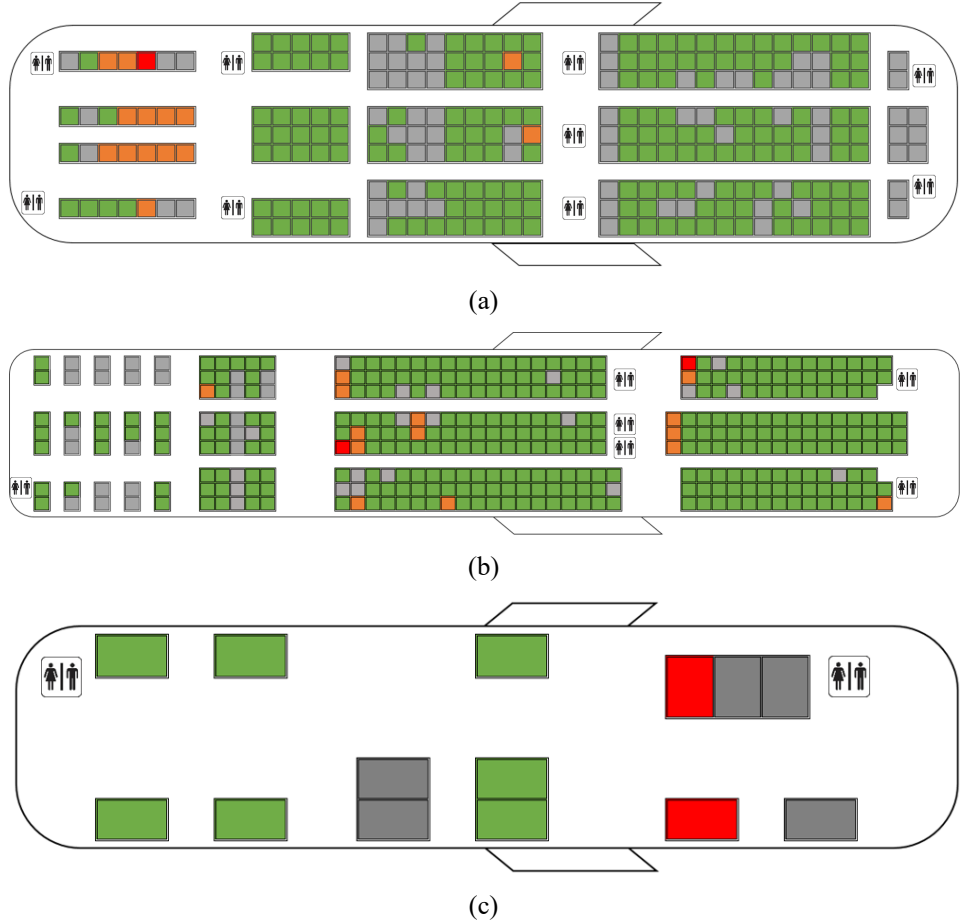

**Figure 1.** Distribution of infections states on flights. Red denotes index cases, orange secondary infections, green PCR negative, and grey empty seats. (a) London flight, (b) Singapore flight, and (c) Japan flight.

## Methods

We use pedestrian dynamics to model the movement of passengers during boarding and deplaning. and inflight movement. We use results from existing literature to model inflight movement [19]. We then input the passenger trajectories and seating arrangements into a fine-scaled infection spread model to identify infection risk. We describe the models and related parameter estimation below.

### *Infection risk model*

The data-driven infection spread model developed here consists of following two components: (i) quantification of the amount of pathogens ingested (infection dose) by a susceptible individual while in close vicinity of an infectious individual, with the dose monotonically decreasing with distance, and (ii) computation of the differential risk of successful transmission during a certain event using infection dose. The primary mechanisms captured in the model include the duration and intensity of the effective contact with an infectious individual and the temporal distribution of distance between individuals, which are obtained from passenger seating and movement during air travel.

The model leverages the results of pedestrian dynamics that captures population mixing behaviors that depend on the environment's layout, the behavioral preferences of people, and proxemic behavior of walking groups, which determine an individual's path of movement. Pedestrian dynamics provides the trajectories of people by outputting the position of each person every Δt seconds. If we know the positions of infective and susceptible persons at a given time, then we can estimate the virus dose that each susceptible person is exposed to in a small time-interval using our new model.

We sum the virus dose over all time intervals to find a normalized, unitless measure $V_n$ of total dose received by the $n^{th}$ susceptible person over the duration of the simulation. There are numerous dose-response relationships available to estimate infection probability from the dose [18, 19]. The commonly used exponential model yields the infection probability $P_n$ for the $n^{th}$ susceptible person as given in Equation (1). $P_n$ is summed over all passengers to yield the *expected number of infections* over the flight.

$$P_n = 1 - \exp(-V_n) \tag{1}$$

The viral load will decay with distance, and we assume a threshold $d_0$ beyond which the viral load is zero. We note that the viral load decreases with distance and that in a short time interval, the exposure is proportional to the time of exposure. We incorporate these insights into a functional form that gives the dose $V_{n,t}$ on susceptible person $n$ from $M$ infective persons in the $t^{th}$ time step as:

$$V_{n,t} = \kappa . \Delta t . \sum_{m=1}^{M} \left(1 - \frac{d_{n,m}}{d_o}\right)^{\alpha}, d_{n,m} < d_o \tag{2}$$

Here, $d_{n,m}$ is the distance between the $n^{th}$ and $m^{th}$ passengers at time $t$. $V_n$ is obtained by summing $V_{n,t}$ over all the time steps. The parameter $\kappa$ is a measure of the dose a person is exposed to per unit time while α controls how quickly the virus concentration drops with distance. The model parameters $\kappa$, α, and $d_o$ are unknown. They are estimated by fitting against a known superspreading event.

### *Scenario to estimate infection model parameters*

We fit parameters to the above model based on a superspreading event in a restaurant in Guangzhou, China early in the pandemic [20]. This situation involves no movement, which makes parameter estimation easier than it would be otherwise. We focus on one room with five occupied tables where one infected person spread the infection to several others belonging to three families. Further details on the parameter estimation are provided in the appendix. Parameter values of $\kappa = 0.15$ minute$^{-1}$, α = 2.5, and $d_0 = 3.5$ m fit the data.

We next explain how we account for a counterfactual situation where people wear masks. Masks act in two ways; first, they reduce the level of contagion in the vicinity of the index case and second, they decrease the distance threshold for viral activity. We vary the parameters $\kappa$ and $d_0$ to account for these two factors. N95/FFP2 masks are roughly 97% effective in preventing leakage under normal fit [21]; therefore, we reduce the $\kappa$ parameter to 3% of the no-mask case. For a regular mask, we use the filtration efficiency of 50% for a cotton bandana "folded surgeon general style" [22]. Another report indicates that the distance traveled by the by respiratory droplets halves at any given time with surgical mask usage [23]. So, we take $d_0$ as 1.7 m for all masks.

*Pedestrian dynamics for boarding*

Social force models for pedestrian dynamics model the pedestrians as particles whose motion is determined by a balance of repulsive and propelling forces [10]. While the agency of the pedestrians to reach a specified target is described in propelling force, the tendency to avoid collision and impenetrability with other individuals in high-density crowds and immobile obstacles in the pedestrian's path are represented by the repulsive terms. These repulsive and attractive forces are summed to obtain the net force acting on $i^{th}$ pedestrian (or particle) as shown in Equation (3), with further details provided in [14].

$$\bar{f}_i = \frac{m_i}{\tau}\left(\bar{v}_o^i(t) - \bar{v}^i(t)\right) + \sum_{j \neq i} \bar{f}_{ij}(t) \qquad (3)$$

The dynamics of pedestrian movement is accomplished by obtaining the velocity and positions at next time steps through numerical integration. In prior work, this model has been validated and applied to movement of people in airplanes and pedestrian queues, where the results from Equation (3) are augmented by human behavior features, such as time for stowing luggage and seat conflicts [11-13]. The pedestrian model parameters are based on our previous study [14]. There is stochasticity as to the order of boarding within each cabin. This is accounted for by averaging the trajectories over 50 simulations.

*Modeling inflight movement*

Hertzberg et al indicate that about 62% of the passengers move from their seat for a median duration of 5.4 minutes (167, in flights of 211-313 minutes duration. Due to the longer duration of the London flight, we consider that all 56 passengers in business and premium economy cabins move for an average of 5.4 minutes out of their seat and 24% of them leave the seat more than once. The typical path of the passenger involves movement to the closest restroom and back to the seat. To account for stochasticity in pedestrian movement, we performed 50 simulations of inflight pedestrian movement for each case analyzed.

Given the location of the index case in the business class cabin, the in-flight movement of passengers in the economy cabin at the back of the aircraft is not relevant, as it would not bring those passengers into contact with the index case. Further, based on equation (1), we can combine the dose due to the different processes (seated co-location, inflight movement and, boarding/deplaning) and combine the resulting probabilities.

**Results**

*Impact of passenger movement*

We estimate infection risk in the London flight in order to (i) verify that the model yields reasonable results in the absence of masks and (i) to examine the role of passenger movement in explaining superspreading patterns. Table 1 shows that the expected number of cases from the simulations – 12.86 – is close to the observed 13 cases. In addition, it also explains the two cases observed in the economy section. Figure 2 illustrates the infection risk probabilities.

We next examine the impact of passenger movement by simulating (i) no passenger movement, (ii) inflight movement without boarding/deplaning, and (iii) only boarding/deplaning. Without passenger movement, the simulations estimate around nine secondary infections in the Business class cabin, which is not far off from the observed eleven cases. But it yields no infections in the premium economy and economy cabins. We next examine if inflight movement can explain these cases. However, adding inflight movement results only in expected 0.4 cases in the premium economy and economy cabins. Table 1 shows that the boarding/deplaning processes contribute more to infection risk than inflight movement does. Note that the combined expected number of cases for the flight is less than for (i) and (iii) combined because the infection from (ii) and (iii) are not mutually exclusive.

**Table 1.** Simulation results for secondary infection by passenger status during the London flight

| Passenger status | Total secondary infections | Secondary infections in business class | Secondary infections in the economy cabins |
|---|---|---|---|
| Only seated | 9.1 | 9.1 | 0 |
| Seated + inflight movement | 9.8 | 9.4 | 0.4 |
| Boarding/deplaning | 4.4 | 3.3 | 1.1 |
| Complete flight | 12.34 | 10.86 | 1.48 |

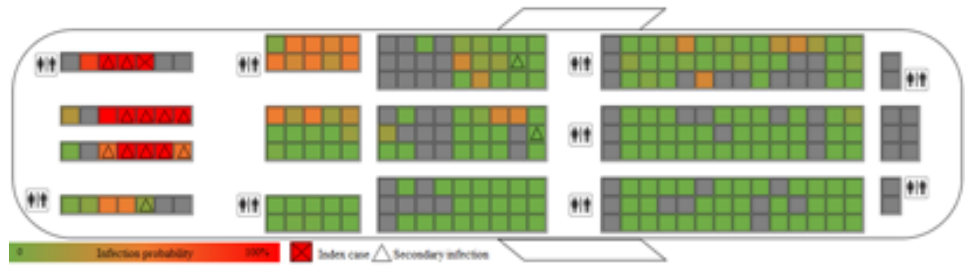

**Figure 2.** Model results of the distribution of secondary infections in the London flight for the duration of the flight.

*Impact of mask usage*

We now examine the impact of different types of masks, using the Singapore flight, where masks were mandated, as the example. The specific type of masks used is not known. We assume a regular mask. The estimated number of cases from the simulations, including passenger movement, is 10.7. This matches well with the 12 secondary cases reported to be from the flight [15]. Figure 3 shows the risk profile as a function of the seat.

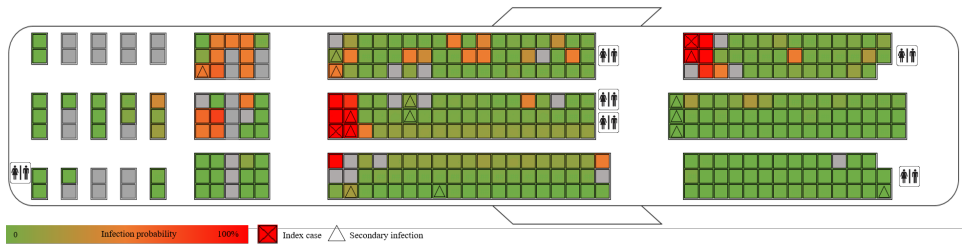

**Figure 3.** Model results of the distribution of secondary infections in the Singapore flight for the duration of the flight.

We now examine the counterfactuals of N95/FFP2 mask use and no mask use. If everyone had used FFP2 or N95 masks for the entire duration of the flight, then the model indicates that there would be 2.3 secondary infections. If there had been no mask usage, then there would be 55 secondary infections.

We next examine the impact of different mask leakages, varying from 3% (N95 with normal fit) to 100% (no mask) to provide insight on the impact of different mask qualities. We also consider various distance thresholds. These would be useful when the actual distance to which droplets and aerosols travel is known for different airflow patterns and masks. Figure 4 presents these results, which could be used to estimate risk when future empirical results identify suitable values for these two parameters.

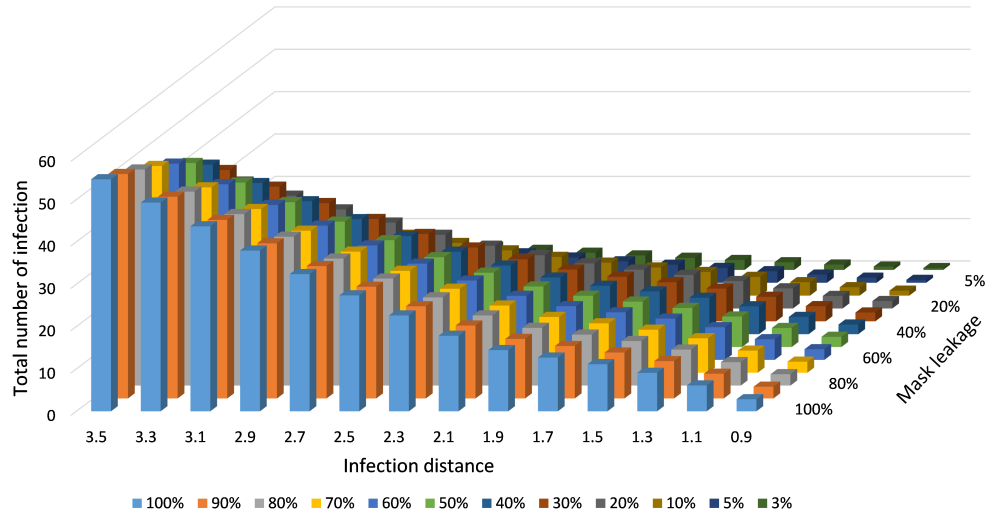

**Figure 4.** Model results of the distribution of secondary infections in the Singapore flight for varying mask leakage and infectivity distance thresholds.

We finally examine if the different infection results from different mask usages (no mask, regular mask, N95/FFP2) are statistically significant. Table 2 presents the mean and 95% confidence intervals for the three masking cases with 50 simulations each and found that none of the upper or lower bounds overlap. Thus, each masking scenario clearly has a different impact.

**Table 2.** Statistical analysis of the impact of masks

| Mask | Middle seat vacant | Mean secondary infections | Upper bound – 95% | Lower bound – 95% |
| --- | --- | --- | --- | --- |
| None | No | 55.03 | 55.30 | 54.76 |
| Cloth | No | 10.46 | 10.48 | 10.44 |
| N95/FFP2 | No | 2.32 | 2.39 | 2.33 |
| None | Yes | 29.75 | 29.93 | 29.58 |
| Cloth | Yes | 5.72 | 5.74 | 5.71 |
| N95/FFP2 | Yes | 0.99 | 0.99 | 0.99 |

### *Validating the impact of N95/FFP2 masks*

To validate the impact of N95/FFP2 masks, we consider the Japan flight, where all passengers used FFP2 masks. This was a long flight (13.5 hours), which we would expect to lead to high infection risk. Simulations with parameters corresponding to FFP2 masks estimate 0.02 new infections with the two index cases in this flight. This is a good estimate of the zero observed cases. Our simulations showed that regular masks would have resulted in 1.7 new infections, while no mask would lead to 2.8 new infections. Thus, N95/FFP2 masks conferred significant benefit.

*Impact of vacant middle seats*

In response to COVID-19, many airlines had adopted strategies to lower density including keeping middle seats vacant [24]. Airlines have discontinued such practices as air travel increased in recent months [25]. A recent study suggested that keeping middle seats empty lowers exposure (dose) significantly [26]. But it did not calculate the infection risk, account for passenger movement, or study the impact of masks.

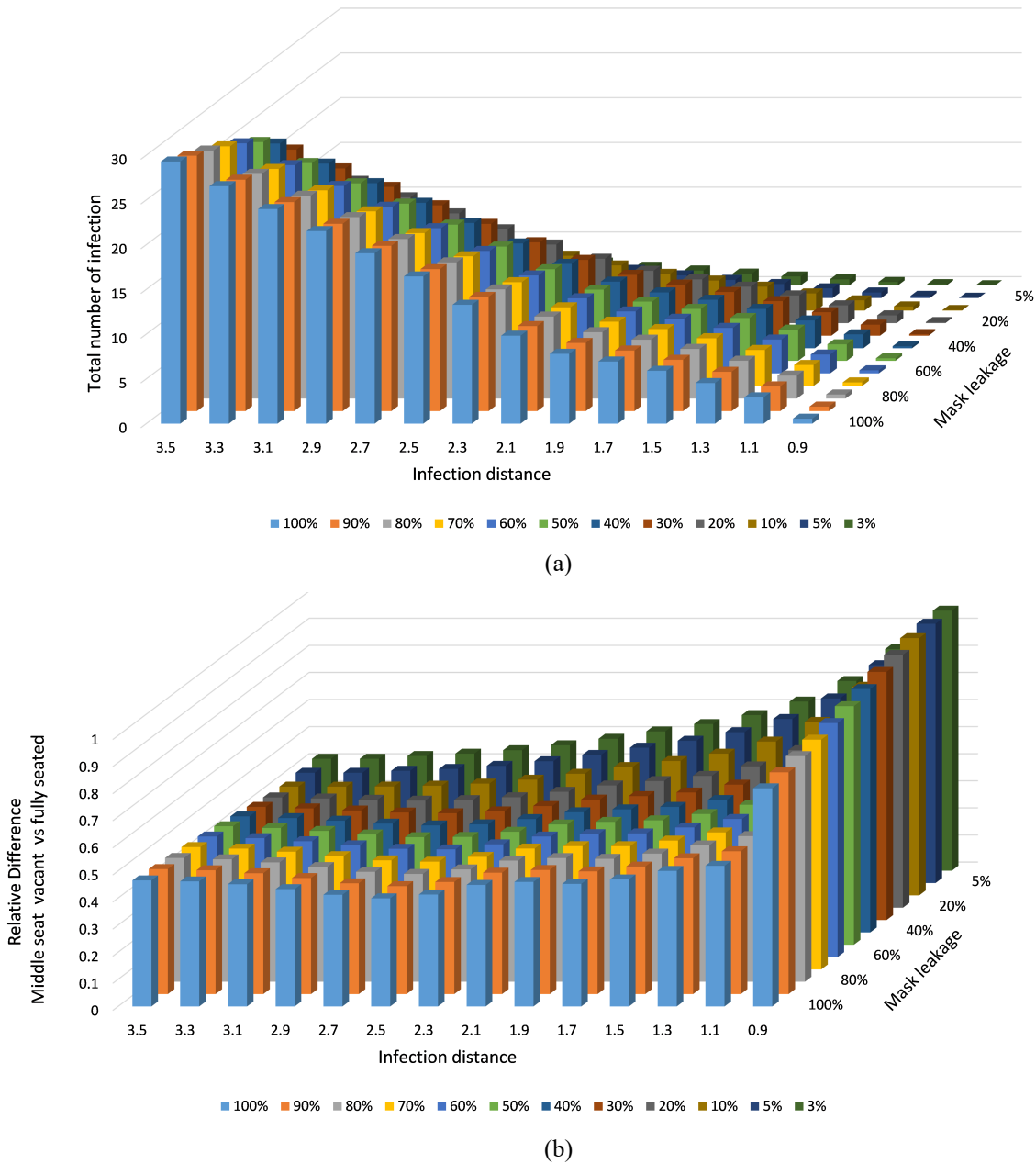

(a)

(b)

**Figure 5.** (a) Model results of the distribution of secondary infections in the Singapore flight for varying mask usage with middle seat vacant. (b) Relative reduction in number of infections comparing middle seat vacant and original aircraft.

Here, we examine the impact of keeping middle seats vacant on infection risk while account for passenger movement. We use the Singapore flight as an example in the results presented in Figure 5. The same index cases, flight duration, boarding, deplaning and inflight movement are considered, but the middle seats in the economy cabin are unoccupied.

*Sensitivity Analysis*

We carry out global parameter sensitivity analysis to evaluate the robustness in model outcomes using the sampling based Partial Rank Correlation Coefficients (PRCC) sensitivity analysis to evaluate variability in model predictions, using the method described in [27, 28]. We examined the effects of the uncertain parameters on the primary outcome the number of new infections, for the Singapore flight with (i) the original configuration, and (ii) with middle seat vacant. PRCCs are used to identify the key parameters contributing to the imprecision in predicting the future infection probability. Details are provided in the appendix.

Our results shown in Figure 6 indicate that both input parameters have a positive PRCC. Both parameters have positive PRCC values and were significantly different from 0 (p-value < 0.05). The results suggest that for both outcome variables, infection distance parameter is most influential in determining the magnitude of outcome variable (|PRCC| > 0.9 at p < 0.05 significance level). However, the influence of both input variables ($\kappa$ reduction and $r_o$) is much higher for the middle seat vacant case compared to the original seating configuration.

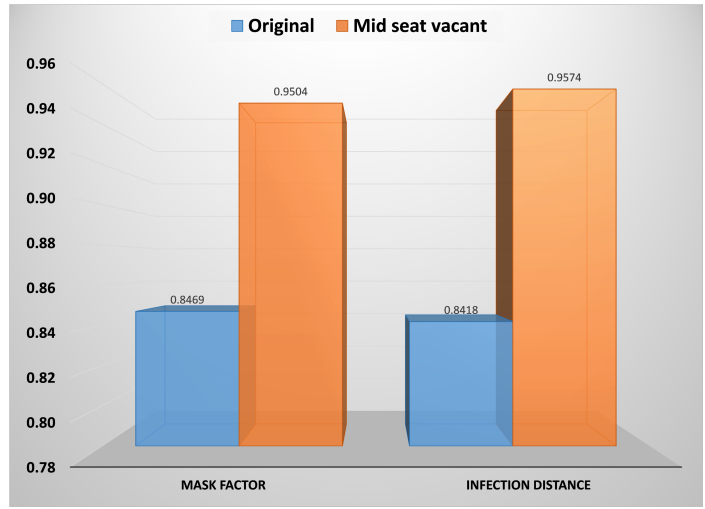

**Figure 6.** PRCC of the number of infections for the original Singapore flight configuration and with middle seat vacant.

**Discussion**

Our method is able to explain the superspreading events on the Hanoi and Singapore flights. Typical simulations and empirical studies do not include movement and so cannot explain such spread [7, 29]. Consequently, contact tracing is performed only for two rows in front and back of the index cases [3]. We show that inclusion of passenger movement can quantitatively explain the spread of infection far from the index cases. Furthermore, boarding/deplaning has a greater impact than inflight movement.

Empirical observation even from before the COVID-19 pandemic had shown that infection can spread far from the index case and that such cases can contribute significantly to the average number of secondary infections during outbreaks on flights [30]. On one flight with distant secondary transmission of COVID-19, the use of a common toilet was suggested as the cause [31], although the general possibility from boarding was also identified. The London flight had separate restrooms for business and economy classes, and so this cannot explain the infection spread while boarding does.

We have also considered the impact of masking. Infection spread has been observed on flights with passengers wearing surgical masks [32], which are more effective than cloth masks. Our results suggest that FFP2 or equivalent masks can almost eliminate all risk of secondary infections during a potential superspreading event, with 95-100% reduction on the flights considered in this paper. Cloth masks are not as effective, although they are considerably better than no mask, leading to 40-80% reduction in cases. Consequently, the number of cases with an N95/FFP2 would be a factor ten lower than with a cloth mask.

Keeping middle seats empty can reduce infection risk by (i) reducing the number of people exposed and (ii) reducing exposure to the contagion through distancing. In the Singapore flight, the number of passengers with this strategy is reduced from 321 to 222, a 30.8% reduction. The reduction in the fraction of infections is greater than this, showing the impact of reduced exposure. However, except for the lowest distance threshold, the reduction in the number of persons plays a greater role in the reduced risk. A reduction in the distance threshold plays a greater role than mask efficiency in reducing infection risk.

Our results suggest the following. (i) Good quality masks ought to be recommended; regular masks have significantly lower impact on long flights. (ii) Leaving middle seat empty is effective and its effectiveness increases when combined with good quality masks. However, with a good mask, the risk is quite low even without middle seats empty. Consequently, use of an N95 mask might be financially more viable than keeping middle seats vacant. (iii) Effective boarding strategies could mitigate the risk of a wide infection outbreak on flights [21, 30]. It is not the dominant factor, but it does play a noticeable role.

Our work has the following limitations. (i) We do not aim to predict the risk of a superspreading event. Rather, we aim to obtain insight into superspreading events so that the risk of such events could be reduced. (ii) We have not accounted for social interactions at the boarding gate, or during other aspects of air travel. (iii) We used empirical comparison against the Japan flight to validate our model with FFP2 masks. However, there is a possibility that the index cases on that flight were not superspreaders. Nevertheless, the relative impacts of the masks would be accurately captured in the simulations. (iv) We have not accounted for vaccinations. The number of secondary infections can be expected to decrease, although the different strategies would have similar relative benefits because the response would still be a monotonic function of the dose. (v) In future work, we intend generalizing the parameter estimation so that results from the mechanisms of infection spread, such as through computational fluid dynamics modeling along with knowledge of virus shedding distributions, could be used to calibrate it for new scenarios.

**Acknowledgments**

This material is based upon work supported by the National Science Foundation under grant numbers 1931511, 2027514, 1931483, and 2027518. Any opinions, findings and conclusions or

**Appendix**

*Distance to the infective person*

The appendix to [19] includes detailed seating arrangements, from which we calculated the distance of each person to the infective person. The superspreader belonged to the family in Table A, and so we exclude that family from the analysis, because they could have been infected through contact elsewhere. We consider other tables, two of which yielded secondary infections – Tables B and C, and two of which did not – Tables E and F. This study also gave the overlap between the infective person and Tables B and C as 53 minutes and 73 minutes for Tables B and C respectively. The number of secondary infections was between four and five, because there was a possibility of one person being exposed to COVID-19 elsewhere. Information on the exposure is provided in Table 3 below.

*Parameter estimation*

We fit parameters to the infection model as follows. There is only one infected person; so, $M = 1$, $r_{n,m}$ is independent of time, giving the distance between the infective person and a susceptible person, $\Delta t$ is taken as the exposure time in Equation (2), and there is only one time step, giving $V_{n,t} = V_n$. We substitute $V_n$ into Equation (1) to get the probability of infection for each susceptible person. We add the infection probabilities of all susceptible persons to get the expected number of susceptible persons. We determine the range of parameters for which the number of infected persons is between 4 and 5. We also required the parameter range to reflect the absence of secondary infections in Tables D and E by yielding the expected number of infections as close to 0.

*Sensitivity analysis procedure*

Following PRCC methodology described in [26, 27], we rank the uncertain parameters, $\kappa$ and $d_0$, in the sampling matrix together with the outcome measures. The PRCC measures the effect of each input parameter on outcome variable, assuming the parameters to be independent. A positive PRCC value indicates that an increase in that parameter leads to an increase in an outcome variable, while a negative value shows that increasing that parameter decreases the outcome variable. Two linear regression models are generated in response to each parameter and outcome measure. A Pearson rank correlation coefficient for the residuals from the two regression models gives the PRCC values for that specific parameter. We consider a uniform distribution for all model parameters. PRCC and p-value of the data are computed using 100 runs of sampling from input parameters distribution.

**Table 3.** Exposure distance and time to infective person.

| Person | Distance (m) | Exposure time (minutes)* | Secondary infection status |
| --- | --- | --- | --- |
| B1 | 1.796 | 53 | Infected |
| B2 | 0.998 | 53 | Infected |
| B3 | 1.531 | 53 | Infected |
| B4 | 2.117 | 53 | Not infected |
| C1 | 2.833 | 73 | Infected |
| C2 | 3.398 | 73 | Possibly infected |
| C3 | 3.182 | 73 | Not infected |
| C4 | 3.414 | 73 | Not infected |
| C5 | 3.057 | 73 | Not infected |
| C6 | 2.546 | 73 | Not infected |
| C7 | 2.210 | 73 | Not infected |
| E1 | 3.675 | 20 | Not infected |
| E2 | 3.145 | 20 | Not infected |
| E3 | 2.566 | 20 | Not infected |
| E4 | 2.114 | 20 | Not infected |
| E5 | 2.105 | 20 | Not infected |
| F1 | 3.986 | 20 | Not infected |
| F2 | 4.694 | 20 | Not infected |
| F3 | 4.700 | 20 | Not infected |
| F4 | 4.175 | 20 | Not infected |
| F5 | 2.984 | 20 | Not infected |

* The exposure time at Tables D and E was not available. We assumed a shorter exposure time of 20 minutes for those tables.